\documentclass[5p,11pt]{elsarticle}

\usepackage{tikz}
\usepackage{microtype}
\usepackage{amssymb}
\usepackage{amsfonts}
\usepackage{amsthm}
\usepackage{amsmath}
\usepackage{dsfont}
\usepackage{mathrsfs}
\usepackage{enumitem}
\usepackage{tensor}
\usepackage{hyperref}
\usepackage{cuted}

\usetikzlibrary{patterns}

\definecolor{darkblue}{rgb}{0.1,0.1,0.7}
\hypersetup{colorlinks,
           linkcolor={darkblue},
           citecolor={darkblue},
           urlcolor={darkblue},
           pdftitle={Gradient Properties of Perturbative Multiscalar RG Flows to Six Loops},
           breaklinks=true,
           pdfdisplaydoctitle
}

\allowdisplaybreaks

\newcommand{\lsp}{\hspace{0.5pt}}
\newcommand{\lnsp}{\hspace{-0.5pt}}
\newcommand{\nc}[3]{\tensor[^{#2}]{#1}{_{#3}}}

\journal{Physics Letters B}

\begin{document}

\begin{frontmatter}

\title{Gradient Properties of Perturbative Multiscalar RG Flows to Six Loops}
\author{William H.\ Pannell}
\ead{william.pannell@kcl.ac.uk}
\author{Andreas Stergiou}
\ead{andreas.stergiou@kcl.ac.uk}
\affiliation{organization={Department of Mathematics, King's College London},
             addressline={Strand}, 
             city={London},
             postcode={WC2R 2LS},
             country={United Kingdom}}

\begin{abstract}
The gradient property of the renormalisation group (RG) flow of multiscalar theories is examined perturbatively in $d=4$ and $d=4-\varepsilon$ dimensions. Such theories undergo RG flows in the space of quartic couplings $\lambda^I$. Starting at five loops, the relevant vector field that determines the physical RG flow is not the beta function traditionally computed in a minimal subtraction scheme in dimensional regularisation, but a suitable modification of it, the $B$ function. It is found that up to five loops the $B$ vector field is gradient, i.e.\ $B^I=G^{IJ}\partial A / \partial\lambda^J$ with $A$ a scalar and $G_{IJ}$ a rank-two symmetric tensor of the couplings. Up to five loops the beta function is also gradient, but it fails to be so at six loops. The conditions under which the $B$ function (and hence the RG flow) is gradient at six loops are specified, but their verification rests on a separate six-loop computation that remains to be performed.
\end{abstract}

\end{frontmatter}

\section{Introduction}
Dynamical systems, governed by differential equations of the form
\begin{equation}\label{eq:dynsys}
    \frac{d\Vec{x}}{dt}=\Vec{F}(\Vec{x})\,,
\end{equation}
appear in every corner of the natural world and are studied in depth in many branches of science and engineering. While general equations like \eqref{eq:dynsys} are often intractable with analytic methods, in some favourable cases features of such systems can be drawn out of the equations themselves, without having to resort to finding explicit solutions. One such case is that of gradient systems, where the function $\Vec{F}(\vec{x})$ may be written in the form
\begin{equation}
    \Vec{F}(\vec{x})=\Vec{\nabla}f(\vec{x})\,.
\end{equation}
Systems undergoing gradient flow have remarkable properties, and it is known from a theorem in \cite{lojasiewicz1982trajectoires} that at late times all solutions must flow to either a fixed point, where $\Vec{F}(\Vec{x})$ vanishes, or off to infinity.

In quantum field theory (QFT), the flow of field theories under a change of energy scale, governed by what is called the renormalisation group (RG), is one such dynamical system. As recognised in \cite{Wilson:1971bg}, the qualitative behaviour of such flows is of immense consequence to the physical system being modelled. In particular, a potential gradient property for the RG flow would have remarkable consequences, e.g.\ it would preclude periodic flow lines and severely limit the nature of potential endpoints of the RG flow.

In this Letter we will be concerned with the gradient property of RG flow in multiscalar QFTs with $N$ real scalar fields in $d=4$ and $d=4-\varepsilon$ dimensions. These QFTs are defined by the Lagrangian
\begin{equation}
    \mathscr{L}=\tfrac12\partial^\mu\phi_i\lsp\partial_\mu\phi_i + \tfrac{1}{4!}\lambda_{ijkl}\phi_i\phi_j\phi_k\phi_l\,,
\end{equation}
where indices from the Latin alphabet take values from 1 to $N$\footnote{We follow the Einstein summation convention for repeated indices. We have assumed a $\mathbb{Z}_2$ symmetry generated by $\phi\to-\phi$ and neglected the allowed term $\tfrac12 \kappa_{ij}\phi_i\phi_j$, which plays no role in our discussion.}. Such theories are renormalisable in perturbation theory. This requires the bare symmetric tensor $\lambda_{ijkl}$ to be substituted with a renormalised one, which we will also call $\lambda_{ijkl}$, which depends on the renormalisation scale $\mu$:
\begin{equation}
    \mu\lsp\frac{d\lambda_{ijkl}}{d\mu}=\beta_{ijkl}\,.
\end{equation}
The quantity $\beta_{ijkl}$, called the beta function, is a function of $\lambda$ and can be computed order by order in a weak-coupling expansion. Its first quantum contribution is of order $\lambda^2$ and takes the form
\begin{equation}\label{eq:beta1}
    \beta_{ijkl}\supset \lambda_{ijmn}\lambda_{mnkl} + \lambda_{ikmn}\lambda_{mnjl} + \lambda_{ilmn}\lambda_{mnjk}\,,
\end{equation}
which follows from the computation of a one-loop Feynman diagram in momentum space and we have rescaled $\lambda$ to eliminate the factor of $1/16\pi^2$ that one encounters. This result can be extended to higher orders, and recently \cite{Bednyakov:2021ojn} used results of \cite{Kompaniets:2017yct} to compute $\beta_{ijkl}$ up to six loops, i.e.\ order $\lambda^7$. Due to the indices carried by $\lambda$ there is a proliferation of contributions to $\beta_{ijkl}$ at higher orders, with 2, 7, 23, 110, 571 distinct terms contributing at two, three, four, five and six loops, respectively.

One may think of $\beta_{ijkl}$ as a vector field $\beta^I$ in the space of couplings, where $I$ denotes the collection of indices $(ijkl)$. A natural question then is if this vector field can be derived as a gradient of a scalar function. In that case the associated flow would be gradient, and the following equation would hold:
\begin{equation}\label{eq:gradient}
    \beta^I = G^{IJ}\partial_J A\,,\quad \partial_I=\partial/\partial\lambda^I\,,
\end{equation}
with $G^{IJ}$ is a rank-two tensor and $A$ a scalar. Both $G^{IJ}$ and $A$ are functions of $\lambda$. The metric $G_{IJ}$ should be Riemannian, i.e.\
\begin{enumerate}[label=(\alph*)]
    \item symmetric, $G_{IJ}=G_{JI}$, and\label{en:metsym}
    \item positive-definite, $G\succ 0$.\label{en:meteigspos}
\end{enumerate}

In the present context of multiscalar theories, the validity of eq.\ \eqref{eq:gradient} with a Riemannian metric was verified up to three loops some time ago in $d=4$ and $d=4-\varepsilon$~\cite{Wallace:1974dx, Wallace:1974dy}, and this was more recently extended to four loops in $d=4$~\cite{Jack:2018oec}.

An important remark is that~\cite{Jack:1990eb, Osborn:1991gm, Jack:2013sha} have already established eq.\ \eqref{eq:gradient} to any order in perturbation theory in $d=4$, but the metric $G_{IJ}$ in their treatment is substituted with a $T_{IJ}$ which is not symmetric in general. However, the symmetry property of the metric has important consequences for potential fixed points of the flow\footnote{Such fixed points lie beyond the computational ability of perturbation theory in $d=4$, but become available in $d=4-\varepsilon$ with $\varepsilon$ small.}. In particular, it is necessary so that around any fixed point local coordinates can be chosen in which the metric is flat to leading order. In that case, $\beta_I=\partial_I A$ at leading order and the matrix $\partial_I\beta_J=\partial_I\partial_J A$ is symmetric at the fixed point. Consequently, its eigenvalues, which determine the scaling dimensions of operators quartic in $\phi$, are real at real fixed points and unitarity is maintained.

Beyond its potential role for unitarity at fixed points of the RG flow, an equation like \eqref{eq:gradient} encapsulates important features of the RG flow itself and has consequences for the existence of possible exotic endpoints one might seek, e.g.\ limit cycles and ergodic behaviour. In short, if eq.\ \eqref{eq:gradient} is valid, then such exotic behaviour is not possible and the quantity $A$ serves as a monotonically decreasing quantity along any RG flow to the infrared, since if $G$ is positive-definite then
\begin{equation}\label{eq:athm}
    \mu\frac{d}{d\mu}A=\beta^I\partial_I A = G_{IJ}\beta^I\beta^J > 0
\end{equation}
for non-zero $\beta^I$. Of course a possible antisymmetric contribution to $G_{IJ}$ does not invalidate \eqref{eq:athm}, so the results of \cite{Jack:1990eb, Osborn:1991gm, Jack:2013sha} are enough to establish that limit cycles and ergodic behaviour are not possible endpoints of RG flows in perturbation theory in $d=4$.

In this work we will consider the gradient property of the RG flow in $d=4$ and we will also describe the extension to $d=4-\varepsilon$. We will analyse the RG flow to six loops, to test the validity of \eqref{eq:gradient} in the most complicated setting available. As we will see, hundreds of non-trivial constraints among beta function coefficients need to be satisfied for the RG flow to be gradient at six loops. The beta function vector field satisfies the required constraints up to five loops, but fails to do so at six loops. However, a suitable modification of it, which gives the proper vector field that describes physical RG flows, may satisfy them but this rests on some hitherto undetermined five- and six-loop results. These can be obtained within the standard paradigm of renormalised perturbation theory, but we do not attempt the associated computations in this work.

\section{Summary of results up to four loops in \texorpdfstring{$d=4$}{d=4}}
Starting with \eqref{eq:beta1} at one loop, it is trivial to verify \eqref{eq:gradient} with
\begin{equation}
    \nc{G}{1}{ijkl;mnpq}=\delta_{ijkl;mnpq}\,,\quad \nc{A}{1\!}{} = \lambda_{ijmn}\lambda_{mnpq}\lambda_{pqkl}\,.
\end{equation}
Our notation $\nc{\!X}{L}{}$ captures the loop order $L$ that the quantity $X$ is first relevant in eq.\ \eqref{eq:gradient}. In a diagrammatic notation, we may write
\begin{equation}\label{eq:beta1diag}
    \tensor[^1]{\lnsp\beta}{}=\begin{tikzpicture}[scale=0.5,baseline=(vert_cent.base)]
        \node (vert_cent) at (0,0) {$\phantom{\cdot}$};
        \draw (0,0) circle (1cm);
        \filldraw (-1,0) circle (2pt);
        \filldraw (1,0) circle (2pt);
        \draw (-1.7,0.7)--(-1,0);
        \draw (-1.7,-0.7)--(-1,0);
        \draw (1.7,0.7)--(1,0);
        \draw (1.7,-0.7)--(1,0);
    \end{tikzpicture}\,,
\end{equation}
\begin{equation}\label{eq:A1diag}
    \nc{G}{1}{}=\begin{tikzpicture}[scale=0.5,baseline=(vert_cent.base)]
        \node (vert_cent) at (0,0) {$\phantom{\cdot}$};
        \draw (-1,1)--(1,1);
        \draw (-1,0.333)--(1,0.333);
        \draw (-1,-0.333)--(1,-0.333);
        \draw (-1,-1)--(1,-1);
    \end{tikzpicture}\,,\qquad \nc{A}{1\!}{}=
    \begin{tikzpicture}[scale=0.5,baseline=(vert_cent.base)]
        \node (vert_cent) at (0,0) {$\phantom{\cdot}$};
        \node (center) at (0,0) {};
        \def\radius{1cm};
        \node[inner sep=0pt] (top) at (90:\radius) {};
        \node[inner sep=0pt] (left) at (210:\radius) {};
        \node[inner sep=0pt] (right) at (330:\radius) {};
        \draw (center) circle[radius=\radius];
        \filldraw (top) circle[radius=2pt];
        \filldraw (left) circle[radius=2pt];
        \filldraw (right) circle[radius=2pt];
        \draw (top) to[out=270, in=30] (left);
        \draw (top) to[out=270, in=150] (right);
        \draw (left) to[out=30, in=150] (right);
    \end{tikzpicture}\,,
\end{equation}
where summations over proper index permutations are understood where necessary. For example, \eqref{eq:beta1diag} contains three terms, as already seen in \eqref{eq:beta1}.

At two loops
\begin{equation}
    \tensor[^2]{\lnsp\beta}{} =
    \tensor[^2]{b}{}\begin{tikzpicture}[scale=0.5,baseline=(vert_cent.base)]
      \node (vert_cent) at (0,0) {$\phantom{\cdot}$};
      \draw (-1,-1)--(1,1);
      \draw (-1,1)--(1,-1);
      \filldraw (0,0) circle[radius=2pt];
      \filldraw (1,1) circle[radius=2pt];
      \filldraw (1,-1) circle[radius=2pt];
      \draw (1,1)--(2,1);
      \draw (1,-1)--(2,-1);
      \draw (1,1) to[out=240, in=120] (1,-1);
      \draw (1,1) to[out=300, in=60] (1,-1);
    \end{tikzpicture}
    + \tensor[^2]{\lnsp c}{}\,
    \begin{tikzpicture}[scale=0.5,baseline=(vert_cent.base)]
      \draw (-1,0)--(1,0);
      \draw (-0.8,1)--(0,0);
      \draw (-0.8,-1)--(0,0);
      \filldraw (0,0) circle[radius=2pt];
      \filldraw (0.7,0) circle[radius=2pt];
      \filldraw (2.3,0) circle[radius=2pt];
      \draw (1,0)--(3,0);
      \draw (0.7,0) to[out=60, in=120] (2.3,0);
      \draw (0.7,0) to[out=-60, in=-120] (2.3,0);
    \end{tikzpicture}
\end{equation}
and
\begin{equation}
    \tensor[^2]{\!A}{} = \tensor[^2]{\lnsp a}{_1}\,
    \begin{tikzpicture}[scale=0.5,baseline=(vert_cent.base)]
      \node (center) at (0,0) {};
      \def\radius{1cm};
      \node[inner sep=0pt] (n1) at (45:\radius) {};
      \node[inner sep=0pt] (n2) at (135:\radius) {};
      \node[inner sep=0pt] (n3) at (225:\radius) {};
      \node[inner sep=0pt] (n4) at (315:\radius) {};
      \draw (center) circle[radius=\radius];
      \filldraw (n1) circle[radius=2pt];
      \filldraw (n2) circle[radius=2pt];
      \filldraw (n3) circle[radius=2pt];
      \filldraw (n4) circle[radius=2pt];
      \draw (n1) to[out=225, in=135] (n4);
      \draw (n1) -- (n4);
      \draw (n2) to[out=315, in=45] (n3);
      \draw (n2) -- (n3);
    \end{tikzpicture}
  + \tensor[^2]{\lnsp a}{_2}\,
  \begin{tikzpicture}[scale=0.5,baseline=(vert_cent.base)]
      \node (center) at (0,0) {};
      \def\radius{1cm};
      \node[inner sep=0pt] (n1) at (0:\radius) {};
      \node[inner sep=0pt] (n2) at (90:\radius) {};
      \node[inner sep=0pt] (n3) at (180:\radius) {};
      \node[inner sep=0pt] (n4) at (270:\radius) {};
      \draw (center) circle[radius=\radius];
      \filldraw (n1) circle[radius=2pt];
      \filldraw (n2) circle[radius=2pt];
      \filldraw (n3) circle[radius=2pt];
      \filldraw (n4) circle[radius=2pt];
      \draw (n1) to[out=180, in=270] (n2);
      \draw (n2) to[out=270, in=0] (n3);
      \draw (n3) to[out=0, in=90] (n4);
      \draw (n4) to[out=90, in=180] (n1);
    \end{tikzpicture}
  + \tensor[^2]{\lnsp a}{_3}\,
  \begin{tikzpicture}[scale=0.5,baseline=(vert_cent.base)]
      \node (center) at (0,0) {};
      \def\radius{1cm};
      \node[inner sep=0pt] (n1) at (0:\radius) {};
      \node[inner sep=0pt] (n2) at (90:\radius) {};
      \node[inner sep=0pt] (n3) at (180:\radius) {};
      \node[inner sep=0pt] (n4) at (270:\radius) {};
      \draw (center) circle[radius=\radius];
      \filldraw (n1) circle[radius=2pt];
      \filldraw (n2) circle[radius=2pt];
      \filldraw (n3) circle[radius=2pt];
      \filldraw (n4) circle[radius=2pt];
      \draw (n1) to[out=150, in=30] (n3);
      \draw (n1) to[out=210, in=-30] (n3);
      \draw (n2) to[out=300, in=60] (n4);
      \draw (n2) to[out=240, in=120] (n4);
    \end{tikzpicture}\,,
    \label{eq:A2loops}
\end{equation}
\begin{equation}
    \tensor[^2]{G}{}= \tensor[^2]{\lnsp g}{} \,
    \begin{tikzpicture}[scale=0.5,baseline=(vert_cent.base)]
        \node (vert_cent) at (0,0) {$\phantom{\cdot}$};
        \draw (-1,1)--(1,0.333);
        \draw (-1,0.333)--(1,1);
        \draw (-1,-0.333)--(1,-0.333);
        \draw (-1,-1)--(1,-1);
        \filldraw (0,0.666) circle[radius=2pt];
    \end{tikzpicture}\,.
    \label{eq:G2loops}
\end{equation}
Demanding $\beta=G\lsp\partial A$ now gives
    \begin{align}
        \tensor[^2]{\lnsp a}{_1}-\tensor[^2]{\lnsp c}{}&=0\,,\label{eq:2loopeq1}\\
        \tensor[^2]{\lnsp a}{_2}+\tfrac14\lsp\tensor[^1]{b}{}\,\tensor[^2]{\lnsp g}{}&=0\,,\label{eq:2loopeq2}\\
        \tensor[^1]{b}{}\,\tensor[^2]{\lnsp g}{}+2\,\tensor[^2]{\lnsp a}{_3}-3\,\tensor[^2]{b}{}&=0\,,\label{eq:2loopeq3}
    \end{align}
where $\tensor[^1]{b}{}$ is the coefficient in the one-loop beta function ($\tensor[^1]{b}{}=1$ from \eqref{eq:beta1diag}). These can be solved for $\tensor[^1]{\lnsp a}{_3},\tensor[^2]{\lnsp a}{_3},\tensor[^3]{\lnsp a}{_3}$ with $\tensor[^2]{\lnsp g}{}=0$ for any $\tensor[^1]{b}{}, \tensor[^2]{b}{}, \tensor[^2]{\lnsp c}{}$, and so there are no constraints among beta function coefficients and the metric can remain globally flat up to this order (although in general $\tensor[^2]{\lnsp g}{}$ is simply an undetermined coefficient at this order, and in fact it remains undetermined even when higher loops are considered below). 

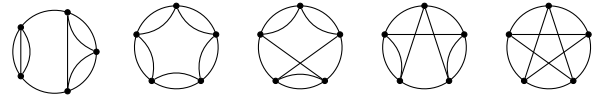
\begin{figure}[ht]
    \centering
\begin{tikzpicture}[scale=0.55,
    vertex/.style={draw,circle,fill=black,minimum size=2pt,inner sep=0pt},
    arc/.style={},
    ]
    \foreach [count=\i] \coord in {(-0.809,0.588),(0.309,0.951),(-0.809,-0.588),(0.309,-0.951),(1.00,0)}{
        \node[vertex] (p\i) at \coord {};
    }
    \draw[arc] (p1)
                    edge (p3)
                    edge [bend left=35] (p3);
    \draw[arc] (p2) edge (p4)
                    edge [bend right=30] (p5);
    \draw[arc] (p5) 
                    edge [bend right=30] (p4);
    \draw circle [radius =1];
\end{tikzpicture}\quad
    \begin{tikzpicture}[scale=0.55, rotate=18,
    vertex/.style={draw,circle,fill=black,minimum size=2pt,inner sep=0pt},
    arc/.style={},
    ]
    \foreach [count=\i] \coord in {(-0.809,0.588),(0.309,0.951),(-0.809,-0.588),(0.309,-0.951),(1.00,0)}{
        \node[vertex] (p\i) at \coord {};
    }
    \draw[arc] (p1)
                    edge [bend right=30] (p2)
                    edge [bend left=30] (p3);
    \draw[arc] (p2)
                    edge [bend right=30] (p5);
    \draw[arc] (p5)
                    edge [bend right=30] (p4);
    \draw[arc] (p4)
                    edge [bend right=30] (p3);
        \draw circle [radius =1];
\end{tikzpicture}\quad
\begin{tikzpicture}[scale=0.55, rotate=162,
    vertex/.style={draw,circle,fill=black,minimum size=2pt,inner sep=0pt},
    arc/.style={},
    ]
    \foreach [count=\i] \coord in {(-0.809,-0.588),(0.309,0.951),(-0.809,0.588),(0.309,-0.951),(1.00,0)}{
        \node[vertex] (p\i) at \coord {};
    }
    \draw[arc]  (p1)    edge    (p2)
                        edge[bend left=30]  (p4);
    \draw[arc]  (p3)    edge    (p5)
                        edge[bend right=25] (p2);
    \draw[arc]  (p5)
                        edge[bend right=30] (p4);
    \draw circle [radius =1];
\end{tikzpicture}\quad
\begin{tikzpicture}[scale=0.55, rotate=162,
    vertex/.style={draw,circle,fill=black,minimum size=2pt,inner sep=0pt},
    arc/.style={},
    ]
    \foreach [count=\i] \coord in {(-0.809,0.588),(0.309,0.951),(0.309,-0.951),(-0.809,-0.588),(1.00,0)}{
        \node[vertex] (p\i) at \coord {};
    }
    \draw[arc]  (p1)
                        edge    (p3)
                        edge[bend left=30] (p4);
    \draw[arc]  (p2)    edge    (p3)
                        edge[bend right=30] (p5);
    \draw[arc]  (p5)    edge    (p4);
    \draw circle [radius =1];
\end{tikzpicture}\quad
\begin{tikzpicture}[scale=0.55, rotate=18,
    vertex/.style={draw,circle,fill=black,minimum size=2pt,inner sep=0pt},
    arc/.style={},
    ]
    \foreach [count=\i] \coord in {(-0.809,0.588),(0.309,0.951),(-0.809,-0.588),(0.309,-0.951),(1.00,0)}{
        \node[vertex] (p\i) at \coord {};
    }
    \draw[arc]  (p1)
                        edge    (p4)
                        edge    (p5);
    \draw[arc]  (p2)
                        edge    (p4)
                        edge    (p3);
    \draw[arc]  (p5)
                        edge    (p3);
    \draw circle [radius =1];
\end{tikzpicture}
\caption{The five vacuum diagrams contributing to $\tensor[^3]{\!A}{}$.}
\label{vacs3}
\end{figure}

At three loops the beta function again has the gradient property regardless of the specific values of coefficients that appear in it, but the metric cannot be chosen to be globally flat any more~\cite{Wallace:1974dy}. That the metric can be globally flat at two loops but not at three loops can be seen by a cursory examination of the diagrams in (\ref{eq:A2loops}) and Fig.~\ref{vacs3}. Each of the diagrams in (\ref{eq:A2loops}) has equivalent vertices, so that there will be a one-to-one correspondence between coefficients in $\tensor[^2]{\beta}{_I}$ and in $\partial_I\lsp\tensor[^2]{\!A}{}$. A cursory examination of Fig.~\ref{vacs3} shows that at three-loops the complexity of the vacuum diagrams increases, erasing this property as a general feature. 
Further, at four loops the gradient property of the beta function vector field is not totally general, and rests on four constraints among beta function coefficients~\cite{Jack:2018oec}. These are indeed satisfied, so the RG flow remains gradient in $d=4$ to four loops.

\section{Diagram generation}
To consider higher orders in the loop expansion,  we need to find efficient ways to generate the terms contributing to $A$ and $G_{IJ}$. To that end, we found it useful to map strings of $\lambda_{ijkl}$ tensors to graphs analogous to Feynman diagrams. Determining whether or not two tensor structures are equivalent up to a permutation of their free indices then becomes a simple application of known algorithms for distinguishing between non-isomorphic graphs. These graph algorithms were handled using the \href{http://szhorvat.net/pelican/igraphm-a-mathematica-interface-for-igraph.html}{\texttt{IGraph/M}} package \cite{Horvát2023} for \textit{Mathematica}, while the tensors themselves were implemented using \href{http://www.xact.es/}{\texttt{xAct}} with \href{http://www.xact.es/xtras/}{\texttt{xTras}} \cite{NUTMA20141719}.

As $A$ is a scalar function, all of the indices on its constituent $\lambda$'s must be contracted, so that the resulting graphs will be the vacuum diagrams in the QFT. As the $n$-th loop order terms in the beta function will be order $\lambda^{n+1}$, the corresponding $n$-loop terms in $A$ will look like
\begin{equation}
    \nc{\!A}{n}{}\supset \sum_{\text{Bubbles}} \nc{\lnsp a}{n}{i}  \;\; \begin{tikzpicture}[baseline=(vert_cent.base)]
        \node (vert_cent) at (0,0) {$\phantom{\cdot}$};
        \draw[pattern=north west lines, pattern color=black] (0,0) circle (0.5cm);
        \filldraw (-0.5,0) circle (1pt);
        \filldraw (-0.4,0.3) circle (1pt);
        \draw[thick, dotted] (-0.35,0.45) arc (130:95:0.45);
        \filldraw (0,0.5) circle (1pt);
    \end{tikzpicture}\,,
\end{equation}
where the sum is over connected bubble diagrams containing $n+2$ vertices and the $\nc{\lnsp a}{n}{i}$ are arbitrary constants. The beta function will only include contributions from connected graphs, so that only connected, one-vertex-irreducible (1VI) graphs can contribute to $A$ with non-zero coefficients. We can then decompose the diagrams in $A$ as
\begin{equation}
    \begin{tikzpicture}[scale=0.5,baseline=(vert_cent.base)]
        \def\radius{1cm};
        \node (vert_cent) at (0,0) {$\phantom{\cdot}$};
        \node[inner sep=0pt] (u1) at (45:\radius) {};
        \node[inner sep=0pt] (l1) at (-45:\radius) {};
        \filldraw (u1) circle (2pt);
        \filldraw (l1) circle (2pt);
        \draw[pattern=north west lines, pattern color=black] (0,0) circle (1cm);
        \draw (0.707,0.707) arc (130:37:0.94);
        \draw (0.707,-0.707) arc (-130:-37:0.94);
        \draw (2.1,0) circle (0.5cm);
        \filldraw (2.1,0.5) circle (2pt);
        \filldraw (2.1,-0.5) circle (2pt);
        \draw (2.1,-0.5)--(2.1,0.5);
    \end{tikzpicture}\,,\qquad
    \begin{tikzpicture}[scale=0.5,baseline=(vert_cent.base)]
        \node (vert_cent) at (0,0) {$\phantom{\cdot}$};
        \draw[pattern=north west lines, pattern color=black] (0,0) circle (1cm);
        \draw (2,0) circle (0.75cm);
        \filldraw (0,1) circle (2pt);
        \filldraw (0,-1) circle (2pt);
        \filldraw (0.707,0.707) circle (2pt);
        \filldraw (0.707,-0.707) circle (2pt);
        \filldraw (2,0.75) circle (2pt);
        \filldraw (2,-0.75) circle (2pt);
        \draw (0,1) arc (130:35:1.4);
        \draw (0.707,0.707)arc(105:75:3);
        \draw (0.707,-0.707)arc(-105:-75:3);
        \draw (0,-1) arc (-130:-35:1.4);
    \end{tikzpicture}\,,\qquad
    \begin{tikzpicture}[scale=0.5,baseline=(vert_cent.base)]
        \node (vert_cent) at (0,0) {$\phantom{\cdot}$};
        \draw[pattern=north west lines, pattern color=black] (0,0) circle (1cm);
        \filldraw (0,1) circle (2pt);
        \filldraw (0,-1) circle (2pt);
        \filldraw (0.707,0.707) circle (2pt);
        \filldraw (0.707,-0.707) circle (2pt);
        \filldraw (0.383,0.924) circle (2pt);
        \filldraw (0.383,-0.924) circle (2pt);
        \draw (0,1) arc (120:82:2.4);
        \draw (0.707,0.707) arc (-80:-30:1.2);
        \draw (0.707,-0.707) arc (80:30:1.2);
        \draw (0,-1) arc (-120:-82:2.4);
        \draw (0.383,0.924)--(1.5,1.3);
        \draw (0.383,-0.924)--(1.5,-1.3);
        \draw (1.5,-1.3) arc (-30:30:2.6);
        \filldraw (1.5,1.3) circle (2pt);
        \filldraw (1.5,-1.3) circle (2pt);
    \end{tikzpicture}\,.
\end{equation}
The essential information in these diagrams will be contained in an order $\lambda^n$ tensor with two, four or six free indices, which we contract with either $\lambda_{iklm}\lambda_{jklm}$, $\lambda_{ijmn}\lambda_{klmn}$, or $\lambda_{ijkp}\lambda_{lmnp}$ respectively.

Similarly, to construct terms in the metric at a given order in $\lambda$ one can notice that they will contain at most three factors of $\delta_{ij}$, so that the $\lambda$-part of the term will have two, four, six or eight external indices. One could also consider constructing the terms in the metric by removing two vertices from a vacuum diagram, so that the connectedness of terms in the beta function requires that we only consider metric terms such that $G^{IJ}\lambda_I\lambda_J$ is a connected, 1VI graph. 

We see that constructing $A$ and $G_{IJ}$ simply requires enumerating all possible combinations of $n$ $\lambda_{ijkl}$ tensors with at most eight free indices. This enumeration is done iteratively, beginning with only $\lambda_{ijkl}$ at $\text{O}(\lambda^1)$ and successively attaching new factors of $\lambda_{ijkl}$ to build up the higher order structures. Depending on the number of contracted indices, attaching a new vertex will change the number of free indices by an integer $m\in\{-4,\,-2,\,0,\,2\}$, leading to the straightforward recursion relations. In order to construct the gradient flow equation up to six loops, we will need the vacuum diagrams to order $\lambda^8$ and the terms in the metric up to order $\lambda^6$. The derivative $\partial A/\partial \lambda_{ijkl}$ corresponds to summing over all of the different ways of removing a vertex, so that the number of equations in (\ref{eq:gradient}) will be equal to the number of these distinct structures. In Table \ref{diagramtable} we list the numbers of the various objects appearing in this paper through six loops.

\begin{table}[ht]
    \centering
    \begin{tabular}{c|c  c  c  c  c  c}
        Loop & 1 & 2 & 3 & 4 & 5 & 6\\
        \hline
        $A$ & 1 & 3 & 5 & 17 & 42 & 177 \\
        $G_{IJ}$ & 1 & 1 & 7 & 18 & 97 & 453\\
        Equations & 1 & 3 & 10 & 36 & 164 & 819
    \end{tabular}
    \caption{Whenever there is overlap, these numbers agree with those listed in \cite{Klebanov:2017nlk,Kleinert:2001hn}.}
    \label{diagramtable}
\end{table}

\section{Results at five and six loops}
At five loops we find 164 equations for 145 unknowns, including those remaining unfixed in $A$ and $G_{IJ}$ at lower orders. One immediately sees that the number of equations has grown faster than the number of unknowns, so that this order even by naive counting we expect that there must exist constraint equations for the coefficients of the various contributions to the beta function. The unknowns $\{\nc{a}{n}{m},\,\nc{g}{n}{m}\}$ do not all appear independently, and we can only solve for 127 of the coefficients, with 18 remaining as free parameters. This leaves 37 of the five-loop equations as constraints on the beta function. We find that these constraints are indeed satisfied by the $\overline{\text{MS}}$ values of the beta function coefficients $\{\nc{b}{n}{m},\,\nc{c}{n}{m}\}$ (where $n\leq 5$).

Many of the constraints are complicated and entirely mysterious. Some, however, may be derived readily from simple arguments. Consider, for instance, the constraint
\begin{equation}
    \nc{b}{5}{67}=\nc{b}{5}{69}\,\left(=36\lsp \zeta_3^2\right)\,,
    \label{eq:constraintex}
\end{equation}
involving the diagrams\footnote{Note that both diagrams in \eqref{eq:constraintgraphs} are primitive (i.e.\ have no subdivergences) and thus the value of the two coefficients given in \eqref{eq:constraintex} is scheme-independent.}
\begin{equation}
    \tensor[^5]{\lnsp\beta}{}\supset\nc{b}{5}{67}\,\begin{tikzpicture}[scale=0.5,baseline=(vert_cent.base)]
      \node (center) at (0,0) {};
      \def\radius{1cm};
      \node[inner sep=0pt] (n1) at (45:\radius) {};
      \node[inner sep=0pt] (o1) at (45:1.5cm) {};
      \node[inner sep=0pt] (n2) at (0:\radius) {};
      \node[inner sep=0pt] (n3) at (135:\radius) {};
      \node[inner sep=0pt] (o2) at (135:1.5cm) {};
      \node[inner sep=0pt] (n4) at (225:\radius) {};
      \node[inner sep=0pt] (o3) at (225:1.5cm) {};
      \node[inner sep=0pt] (n5) at (270:\radius) {};
      \node[inner sep=0pt] (n6) at (315:\radius) {};
      \node[inner sep=0pt] (o4) at (315:1.5cm) {};
      \draw (center) circle[radius=\radius];
      \filldraw (n1) circle[radius=2pt];
      \filldraw (n3) circle[radius=2pt];
      \filldraw (n4) circle[radius=2pt];
      \filldraw (n5) circle[radius=2pt];
      \filldraw (n6) circle[radius=2pt];
      \draw (n1) -- (o1);
      \draw (n1) -- (n5);
      \draw (n3) -- (o2);                
      \draw (n3) -- (n6);
      \draw (n4) -- (o3);
      \draw (n6) -- (o4);
      \draw (n4) -- (n2);
      \filldraw (n2) circle[radius=2pt];
      \draw (n2) -- (n5);
    \end{tikzpicture}
    \lsp+\nc{b}{5}{69}\,\begin{tikzpicture}[scale=0.5,baseline=(vert_cent.base)]
      \node (center) at (0,0) {};
      \def\radius{1cm};
      \node[inner sep=0pt] (n1) at (45:\radius) {};
      \node[inner sep=0pt] (o1) at (45:1.5cm) {};
      \node[inner sep=0pt] (n2) at (90:\radius) {};
      \node[inner sep=0pt] (n3) at (135:\radius) {};
      \node[inner sep=0pt] (o2) at (135:1.5cm) {};
      \node[inner sep=0pt] (n4) at (225:\radius) {};
      \node[inner sep=0pt] (o3) at (225:1.5cm) {};
      \node[inner sep=0pt] (n5) at (270:\radius) {};
      \node[inner sep=0pt] (n6) at (315:\radius) {};
      \node[inner sep=0pt] (o4) at (315:1.5cm) {};
      \draw (center) circle[radius=\radius];
      \filldraw (n1) circle[radius=2pt];
      \filldraw (n2) circle[radius=2pt];
      \filldraw (n3) circle[radius=2pt];
      \filldraw (n4) circle[radius=2pt];
      \filldraw (n5) circle[radius=2pt];
      \filldraw (n6) circle[radius=2pt];
      \draw (n1) -- (o1);
      \draw (n1) -- (n5);
      \draw (n3) -- (o2);                
      \draw (n3) -- (n5);
      \draw (n4) -- (o3);
      \draw (n4) -- (n2);
      \draw (n6) -- (o4);
      \draw (n6) -- (n2);
    \end{tikzpicture}\,.
    \label{eq:constraintgraphs}
\end{equation}
As these diagrams are primitive, gradient flow can only hold if they are absorbed by the same term in $\tensor[^5]{\!A}{}$, namely
\begin{equation}
\tensor[^5]{\! A}{}\supset\nc{a}{5}{42}\,\begin{tikzpicture}[scale=0.5,baseline=(vert_cent.base)]
      \node (center) at (0,0) {};
      \def\radius{1cm};
      \node[inner sep=0pt] (c1) at (0,0) {};
      \node[inner sep=0pt] (n1) at (45:\radius) {};
      \node[inner sep=0pt] (n2) at (90:\radius) {};
      \node[inner sep=0pt] (n3) at (135:\radius) {};
      \node[inner sep=0pt] (n4) at (225:\radius) {};
      \node[inner sep=0pt] (n5) at (270:\radius) {};
      \node[inner sep=0pt] (n6) at (315:\radius) {};
      \draw (center) circle[radius=\radius];
      \filldraw (n1) circle[radius=2pt];
      \filldraw (n2) circle[radius=2pt];
      \filldraw (n3) circle[radius=2pt];
      \filldraw (n4) circle[radius=2pt];
      \filldraw (n5) circle[radius=2pt];
      \filldraw (n6) circle[radius=2pt];
      \filldraw (c1) circle[radius=2pt];
      \draw (n1) -- (c1);
      \draw (n1) -- (n5);
      \draw (n2) -- (n4);
      \draw (n2) -- (n6);
      \draw (n3) -- (c1);                
      \draw (n3) -- (n5);
      \draw (n4) -- (c1);
      \draw (n6) -- (c1);
    \end{tikzpicture}\,.
\end{equation}
The constraint \eqref{eq:constraintex} then just expresses the fact that this part of $\tensor[^5]{\lnsp\beta}{}$ must be derivative.

At six loops we find 819 equations for 648 unknowns. Solving these equations for $A$ and $G_{IJ}$, we find a total of 234 constraints at this order\footnote{When considering the constraints in a general scheme we find 237. However, three of these are identically zero in $\overline{\text{MS}}$ as they involve only graphs which are either one-vertex reducible or have non-overlapping divergences.}. Of these, five are not satisfied by the coefficients in $\overline{\text{MS}}$, indicating that $\beta_{ijkl}$ is \emph{not} gradient for a generic scalar field theory at six loops. 

\begin{figure}[ht]
\centering
\begin{tikzpicture}[scale=0.55,baseline=(vert_cent.base)]
    \node (vert_cent) at (0,0) {$\phantom{\cdot}$};
    \node[inner sep=0pt] (center) at (0,0) {};
    \def\radius{1cm};
    \node[inner sep=0pt] (n1) at (45:\radius) {};
    \node[inner sep=0pt] (n2) at (90:\radius) {};
    \draw (center) circle (1cm);
    \draw (-1,0) to[out=0, in=270] (n2);
    \draw (n2)--(center);
    \draw (center)--(1,0);
    \draw (n1) to[out=200, in=70] (center);
    \draw (n1) to[out=250, in=20] (center);
    \draw (-1.7,0)--(-1,0);
    \draw (1,0)--(1.7,0);
    \filldraw (-1,0) circle (1.8pt);
    \filldraw (1,0) circle (1.8pt);
    \filldraw (0,1) circle (1.8pt);
    \filldraw (n1) circle (1.8pt);
    \filldraw (n2) circle (1.8pt);
    \filldraw (center) circle (1.8pt);
  \end{tikzpicture}
  \begin{tikzpicture}[scale=0.55,baseline=(vert_cent.base)]
    \node (vert_cent) at (0,0) {$\phantom{\cdot}$};
    \node (center) at (0,0) {};
    \def\radius{1cm};
    \node[inner sep=0pt] (n1) at (105:\radius) {};
    \node[inner sep=0pt] (n2) at (165:\radius) {};
    \draw (center) circle (1cm);
    \draw (n1)--(n2);
    \draw (n2) to[out=12, in=258] (n1);
    \draw (-1.7,0)--(-1,0);
    \draw (1,0)--(1.7,0);
    \filldraw (-1,0) circle (1.8pt);
    \filldraw (1,0) circle (1.8pt);
    \filldraw (0,1) circle (1.8pt);
    \filldraw (n1) circle (1.8pt);
    \filldraw (n2) circle (1.8pt);
    \draw (-1,0) to[out=0, in=270] (0,1);
    \draw (0,1) to[out=270, in=180] (1,0);
  \end{tikzpicture}
  \begin{tikzpicture}[scale=0.55,baseline=(vert_cent.base)]
    \node (vert_cent) at (0,0) {$\phantom{\cdot}$};
    \node[inner sep=0pt] (center) at (0,0) {};
    \def\radius{1cm};
    \node[inner sep=0pt] (n1) at (270:\radius) {};
    \node[inner sep=0pt] (n2) at (90:\radius) {};
    \draw (center) circle (1cm);
    \draw (-1,0) to[out=0, in=270] (n2);
    \draw (n2)--(center);
    \draw (center)--(1,0);
    \draw (n1) to[out=120, in=240] (center);
    \draw (n1) to[out=60, in=300] (center);
    \draw (-1.7,0)--(-1,0);
    \draw (1,0)--(1.7,0);
    \filldraw (-1,0) circle (1.8pt);
    \filldraw (1,0) circle (1.8pt);
    \filldraw (0,1) circle (1.8pt);
    \filldraw (n1) circle (1.8pt);
    \filldraw (n2) circle (1.8pt);
    \filldraw (center) circle (1.8pt);
  \end{tikzpicture}
  \begin{tikzpicture}[scale=0.55,baseline=(vert_cent.base)]
    \node (vert_cent) at (0,0) {$\phantom{\cdot}$};
    \node (center) at (0,0) {};
    \def\radius{1cm};
    \node[inner sep=0pt] (n1) at (135:\radius) {};
    \node[inner sep=0pt] (n2) at (90:\radius) {};
    \node[inner sep=0pt] (n3) at (270:\radius) {};
    \draw (center) circle (1cm);
    \draw (-1,0) to[out=35, in=270] (n1);
    \draw (n1) to[out=0, in=225] (n2);
    \draw (n3) to[out=90, in=180] (1,0);
    \draw (n2)--(n3);
    \draw (-1.7,0)--(-1,0);
    \draw (1,0)--(1.7,0);
    \filldraw (-1,0) circle (1.8pt);
    \filldraw (1,0) circle (1.8pt);
    \filldraw (0,1) circle (1.8pt);
    \filldraw (n1) circle (1.8pt);
    \filldraw (n2) circle (1.8pt);
    \filldraw (n3) circle (1.8pt);
  \end{tikzpicture}
\caption{The four asymmetrical propagator corrections contributing to $S_{ij}$ at five-loops.}
\label{fig:S5loops}
\end{figure}
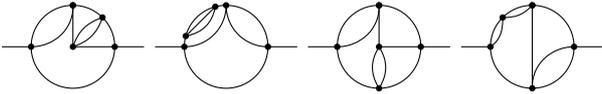

However, beginning at five loops there exists an essential ambiguity in the definition of $\beta_{ijkl}$, which prevents it from being the physical object associated with RG flow \cite{Jack:1990eb, Fortin:2012hn}. This ambiguity arises from a mismatch between the object used to eliminate poles in propagator diagrams, $Z_{ij}$, and the quantity used to renormalise the bare field operator, $Z^{1/2}_{ij}$. While $Z_{ij}$ is used for propagator diagrams, it is $Z^{1/2}_{ij}$ which is used to renormalise the two-point function. This may be calculated from $Z_{ij}$ using the matrix equation
\begin{equation}
    Z=(Z^{1/2})^T\lsp Z^{1/2}\,.
\end{equation}
The crucial observation is that this does not determine $Z^{1/2}$ uniquely, due to the possible $O(N)$ rotation
\begin{equation}
    Z^{1/2}_{ij}\rightarrow R_{ik}\lsp Z^{1/2}_{kj}\,,\qquad R\in O(N)\,,
    \label{eq:ambiguity}
\end{equation}
which does not alter the pole structure of the propagator. If $R$ contains $1/\varepsilon$ poles itself, the beta function can be affected, however. To see this, we will work with an infinitesimal rotation, and expand the Lie algebra element as a series in $1/\varepsilon$,
\begin{equation}
    R_{ij}=\delta_{ij}+\sum_n \frac{O^{(n)}_{ij}}{\varepsilon^n}\,,
\end{equation}
where we have used the notation of \cite{Fortin:2012hn} and denoted the Lie algebra element by $O_{ij}$. One can then see that this will alter the poles in $Z^{1/2}_{ij}$, specifically sending the $1/\varepsilon$ pole to
\begin{equation}
    Z^{1/2,\,(1)}_{ij}\rightarrow Z^{1/2,\,(1)}_{ij}+O^{(1)}_{ij}\,.
\end{equation}
If $O^{(1)}_{ij}$ has dependence on $\lambda_{ijkl}$ this extra term will contribute to both the anomalous dimension matrix $\gamma_{ij}$ and thus also to the beta function, shifting them to become
\begin{equation}
\begin{aligned}
    \gamma_{ij}&\rightarrow\gamma_{ij}-\omega_{ij}\,,\\\ \beta_{ijkl}&\rightarrow\beta_{ijkl}+\left(\omega_{im}\lambda_{mjkl}+\text{perms.}\right)\,,
\end{aligned}
\end{equation}
where $\omega_{ij}=-\lambda_I\partial_I O^{(1)}_{ij}$. This ambiguity prevents the beta function from being physically meaningful as long as we can write down such a correction $O^{(1)}=O^{(1)}(\lambda)$. As $O^{(1)}_{ij}=O^{(1)}_{[ij]}$, these come from asymmetric propagator diagrams which appear first at five loops with the four shown in Fig.\ \ref{fig:S5loops}.
Thus, beginning at five loops there is an essential ambiguity in $\beta_{ijkl}$ which prevents it from being the physical object associated with RG flow.

To resolve this, we must turn to the trace of the stress-tensor. As shown in \cite{Jack:1990eb}, $T^\mu{\!}_\mu$ will not be proportional $\beta_{ijkl}\hat{\mathcal{O}}_{ijkl}$, but rather $B_{ijkl}\hat{\mathcal{O}}_{ijkl}$, where the $B$-function is a modification of the beta function:
\begin{equation}
    B_{ijkl}=\beta_{ijkl}-(S_{im}\lambda_{mjkl}+\text{perms.})\,,\qquad S_{ij}=S_{[ij]}\,.
\end{equation}
The only features of the additional term $S_{ij}$, which can be calculated using counterterms arising from coordinate-dependent couplings, which are needed for this work are that under (\ref{eq:ambiguity}) it will transform as
\begin{equation}
    S_{ij}\rightarrow S_{ij}+\omega_{ij}\,,\qquad
\end{equation}
and that it will have some expansion
\begin{equation}
    S_{ij}=\sum_{n\geq5}\sum_m\nc{s}{n}{m}\nc{\mathcal{S}}{n}{m,\,ij}
\end{equation}
where $\nc{\mathcal{S}}{n}{m,\,ij}$ are anti-symmetrised propagator diagrams, e.g.\ those contained in Fig.\ \ref{fig:S5loops} for five loops. While the leading coefficients $\nc{s}{n}{m}$ have been calculated in the context of scalar-fermion theories in $d=4$ \cite{Fortin:2012hn}, where the first contributions to $S$ appear at three loops, no such computation has been performed for multiscalar theories in $d=4$, and we will leave them unfixed throughout our analysis. From the first property, one immediately sees that, unlike the beta function, $B_{ijkl}$ is invariant under (\ref{eq:ambiguity}) and is thus unambiguous. Beginning at five loops we are then forced to identify the $B$-function as the proper vector field associated with RG flow\footnote{Some renormalisation aspects associated with the shift $\beta\to B$ are discussed in \cite{Herren:2021yur}.}.

Practically, the distinction between $B_{ijkl}$ and $\beta_{ijkl}$ is an antisymmetric shift of coefficients associated with asymmetric propagator diagrams. There are four such diagrams at five loops and 19 at six loops, leading to a total of 23 unknowns which may appear in the constraint equations. For five loops, we define the $\nc{s}{5}{m}$ by using the ordering shown in Fig.\ \ref{fig:S5loops} for $m$, with the left leg getting the index $i$. Implementing this shift into the existing constraints for $\nc{b}{n}{m}$ and $\nc{c}{n}{m}$, we find that at five loops the $\nc{c}{5}{m}$ coefficients associated with Fig.\ \ref{fig:S5loops} appear only symmetrically, so that the factors of $\nc{s}{5}{m}$ cancel. The gradient property of $\beta_{ijkl}$ thus implies the gradient property of $B_{ijkl}$ at five loops. At six loops we find that the five unsatisfied constraints become two constraints on $\nc{s}{5}{m}$. However, four of the previously satisfied constraints also receive contributions from $\nc{s}{n}{m}$, leading to a total of four constraint equations:
\begin{equation}\label{eq:Sconstraints}
\begin{aligned}
\nc{\lnsp s}{5}{2}=\frac{259}{4608}&\,,\qquad 192\,\nc{\lnsp s}{5}{3}+384\,\nc{\lnsp s}{5}{4}=31-36\lsp\zeta_3\,, \\
\nc{\lnsp s}{5}{1}-12\,\nc{\lnsp s}{5}{2}&+\nc{\lnsp s}{5}{4}=12\lsp\nc{\lnsp s}{6}{8}-6\lsp\nc{\lnsp s}{6}{9}-6\,\nc{\lnsp s}{6}{16}\,, \\
4\,\nc{\lnsp s}{5}{1}-2\,\nc{\lnsp s}{5}{3}&=-2\,\nc{\lnsp s}{6}{3}+\nc{\lnsp s}{6}{6}+4\,\nc{\lnsp s}{6}{10}-2\,\nc{\lnsp s}{6}{15}+\nc{\lnsp s}{6}{17}\,.
\end{aligned}
\end{equation}
As these coefficients have not been calculated, these equations amount to a prediction of their values, an explicit one in the case of $\nc{s}{5}{2}$, if the $B$-function is to be gradient. Given that $B_{ijkl}$ already satisfies 266 constraints through six loops, evidence suggests that gradient property is an intrinsic property of multiscalar RG flows, and thus that an explicit calculation of $S_{ij}$ must uphold (\ref{eq:Sconstraints}).

\section{Scheme independence of results}
Up to this point the discussion has been entirely confined to the $\overline{\text{MS}}$ renormalisation scheme. To ensure that our results are physically meaningful statements, we must demonstrate that the constraints are satisfied in all renormalisation schemes. While we considered arbitrary coefficients in deriving the constraints on the beta function, we had implicitly demanded the scheme-dependent statement that the coefficients of tensor structures not appearing in $\overline{\text{MS}}$ were zero. While it is possible to prove scheme-independence by recasting the constraints in terms of explicitly scheme-independent quantities as was done for the four-loop constraints in \cite{Jack:2018oec}, we will take a more brute-force approach.

A change in renormalisation scheme can be realised as a generic shift in the interaction tensor, 
\begin{equation}\label{eq:lamchange}
    \lambda^{I}\rightarrow \lambda'^{\,I}=\lambda^{I}+\sum_{n=1}^\infty\sum_m\nc{r}{n}{m} \tensor[^n]{T}{_{m}}^I(\lambda)\,,
\end{equation}
where the coefficients $\nc{r}{n}{m}$ are arbitrary, and $m$ indexes a sum over distinct tensor structures $\nc{T}{n}{m}$ of order $\lambda^n$. The number of these structures will be identical with the number of equations listed in Table \ref{diagramtable}. In order to determine how this shift will alter the coefficients appearing in the beta function, we must notice that
\begin{align}
    \beta'^{\,I}(\lambda')&=\mu\frac{d\lambda'^{\,I}}{d\mu}\nonumber\\&=\beta^I(\lambda)+\sum_{n=1}^\infty\sum_m\sum_{\hat{T}\subset T}\nc{r}{n}{m} \lsp (\tensor[^n]{\hat{T}}{_{m}})^I_{\,J} \lsp\beta^J(\lambda)\,,
    \label{eq:schemechange}
\end{align}
where $\hat{T}$ has a vertex removed compared to $T$. We also notice that $\beta'_I(\lambda')$ has an expansion in its own right as
\begin{equation}
    \beta'^{\,I}(\lambda')=\sum_{n=1}^\infty\sum_m \nc{\mathfrak{b}}{n}{m} \lsp\nc{T}{n}{m}^I(\lambda')\,,
    \label{eq:betaprimeexpansion}
\end{equation}
where now no $\nc{\mathfrak{b}}{n}{m}$'s are assumed to be zero from the beginning, in contrast with the $\overline{\text{MS}}$ treatment so far. We then notice that we can isolate the new coefficients $\nc{\mathfrak{b}}{n}{m}$ by expanding the $\lambda'^{\,I}$ tensors in \eqref{eq:betaprimeexpansion} using \eqref{eq:lamchange} and then matching with the right-hand side of \eqref{eq:schemechange}. We perform this matching with the same computational method used to find the solution to the gradient flow equations, and find the resulting generic scheme change up to six loops.

Solving the gradient flow equations with $\beta'^{\,I}(\lambda')$, we find again the 275 constraints through six loops, now in terms of the $\nc{\mathfrak{b}}{n}{m}$. At four and five-loops, substituting in the explicit expressions $\nc{\mathfrak{b}}{n}{m}=\nc{\mathfrak{b}}{n}{m}(r,b,c)$ yields constraints that are independent of the scheme-change parameters $\nc{r}{n}{m}$, and reduce to those found in $\overline{\text{MS}}$. At six loops the constraints contain explicit factors of $\nc{r}{n}{m}$, which naively seem to spoil scheme-invariance. However, these terms appear multiplied with combinations of $\tensor[^n]{b}{}$ and $\tensor[^n]{\lnsp c}{}$ beta-function coefficients that vanish upon taking their $\overline{\text{MS}}$ values. These factors will then not spoil the conclusion that satisfying the $\overline{\text{MS}}$ constraints implies that the constraints are satisfied in a general scheme. The gradient property of the beta function in the $\overline{\text{MS}}$ scheme thus ensures the gradient property of the beta function in all schemes.

\section{Extension to \texorpdfstring{$d=4-\varepsilon$}{d=4-epsilon}}
Working slightly away from four dimensions in $d=4-\varepsilon$, the coupling $\lambda_{ijkl}$ becomes weakly relevant and $\beta_{ijkl}$ receives a contribution from its classical scaling dimension in the form of a linear term,
\begin{equation}
    \beta_{ijkl}\supset -\varepsilon\lsp\lambda_{ijkl}\,.
\end{equation}
This term can only be absorbed by introducing a new term in $A$,
\begin{equation}
    \tensor[^0]{\!A}{}=-\frac{\varepsilon}{2}\,\begin{tikzpicture}[scale=0.5,baseline=(vert_cent.base)]
        \node (vert_cent) at (0,0) {$\phantom{\cdot}$};
        \draw (0,0) circle (1cm);
        \filldraw (-1,0) circle (2pt);
        \filldraw (1,0) circle (2pt);
        \draw (1,0) to[out=135, in=45] (-1,0);
        \draw (1,0) to[out=225, in=315] (-1,0);
    \end{tikzpicture}\,.
    \label{eq:epsilonA}
\end{equation}
At $n$ loops this contribution to $A$ will bring in terms in the metric which in four dimensions first appeared at $n+1$ loops, namely
\begin{equation}
    -\varepsilon\sum_m \nc{g}{n+1}{m} \nc{G}{n+1}{m}^{IJ}\lambda_J\,.
    \label{eq:additionalterm}
\end{equation}
Extending our four-dimensional solution to $d=4-\varepsilon$ without the introduction of non-perturbative $1/\varepsilon$ poles\footnote{One notices that a solution always exists at these orders if we permit $1/\varepsilon$ poles to appear. In fact this is fully general, and one can prove that such a solution will exist at all loop orders as $G^{IJ}\lambda_J$ will be a full-rank matrix equation in the coefficients $g$. While the coupling $\lambda_{ijkl}$ will be $\text{O}(\varepsilon)$ at the fixed points, there is nothing preventing a generic perturbative flow from moving beyond this region.} requires that we absorb this new term into $\varepsilon$ corrections to $a$ and $g$:
\begin{equation}
    \nc{a}{n}{m}\rightarrow\nc{a}{n}{m}+\varepsilon\lsp\nc{d}{n}{m}(\varepsilon)\,,\qquad \nc{g}{n}{m}\rightarrow\nc{g}{n}{m}+\varepsilon\lsp\nc{h}{n}{m}(\varepsilon)\,.
\end{equation}
Here $\nc{d}{n}{m}(\varepsilon)$ and $\nc{h}{n}{m}(\varepsilon)$ are arbitrary polynomials in $\varepsilon$. These $\varepsilon$ corrections will alter the gradient flow equation to read, schematically,
\begin{equation}
    \varepsilon\left(\lambda+H(\varepsilon)\lsp\partial A+G\lsp\partial D(\varepsilon)\right)+\varepsilon^2\lsp H(\varepsilon)\lsp\partial D(\varepsilon)=0\,,
    \label{eq:epsiloneqn}
\end{equation}
where we have discarded the $\text{O}(\varepsilon^0)$ terms already solved for in terms of $a$ and $g$. At $\text{O}(\lambda^6)$ and below the values of $G$ and $A$ will largely be fixed by the four-dimensional solution, but the $\text{O}(\lambda^7)$ equation will contain the novel metric term $\tensor[^7]{G}{}$. As we are truncating the expansion, and have already implicitly discarded all terms in (\ref{eq:gradient}) of $\text{O}(\lambda^8)$ and above, we are free to include the coefficients $\nc{g}{7}{m}$ in the solution at this order. While treating these as free parameters here may not allow for a consistent solution at seven loops, it will be sufficient to draw conclusions through six loops. In general, solutions to (\ref{eq:epsiloneqn}) will include $1/\varepsilon$ poles, which both obstruct the smooth $\varepsilon\rightarrow0$ limit and invalidate the existing $\text{O}(\varepsilon^0)$ solution. Demanding that these poles vanish provides additional equations that $H(\varepsilon)$ and $D(\varepsilon)$ must satisfy.

We solve for $\{\nc{d}{n}{m},\,\nc{h}{n}{m}\}$, and the freedom in $G$ and $A$ unfixed by the four dimensional solution using the same techniques as before. When the dust settles, we find four constraints on the coefficients in the beta function at five loops, and 32 constraints at six loops. The additional freedom in $\tensor[^7]{G}{}$ means that at six loops the constraints only appear from the $\text{O}(\lambda^6)$ equations. We note that these constraints are independent of those found in $d=4$, and crucially are all satisfied in $\overline{\text{MS}}$. Importantly, the coefficients $\nc{s}{n}{m}$ appearing in $S_{ij}$ identically cancel in all of these constraints, so that there is no obstruction to $B_{ijkl}$ satisfying them, still subject to \eqref{eq:Sconstraints}. We exhibit the two loop solution including $\varepsilon$-dependence in \ref{ExplicitAppendix}, with all of the remaining unfixed coefficients $\{\nc{a}{n}{m},\,\nc{g}{n}{m},\,\nc{d}{n}{m},\,\nc{h}{n}{m}\}$ set to zero for convenience.

To establish the physical relevance of the constraints in $d=4-\varepsilon$, we once again examine them under a generic scheme change, replacing the $\nc{\mathfrak{b}}{n}{m}$ in (\ref{eq:betaprimeexpansion}) with
\begin{equation}
    \nc{\mathfrak{b}}{n}{m}\rightarrow \nc{\mathfrak{b}}{n}{m}+\varepsilon\,\nc{\mathfrak{e}}{n}{m}\,.
\end{equation}
Dividing (\ref{eq:schemechange}) into terms of $\text{O}(\varepsilon^0)$ and terms of $\text{O}(\varepsilon)$, we see that the $\nc{\mathfrak{b}}{n}{m}$ will be the same as in four dimensions. The $\text{O}(\varepsilon)$ terms will be
\begin{equation}
    \sum_{n=1}^\infty\sum_m\nc{\mathfrak{e}}{n}{m}\nc{T}{n}{m}^I(\lambda')=-\sum_{n=1}^\infty\sum_m(n-1)\lsp\nc{r}{n}{m} \lsp \tensor[^n]{T}{_{m}}^I(\lambda)\,,
\end{equation}
where we have used the fact that $\nc{\mathfrak{e}}{0}{}=-1$ and (\ref{eq:lamchange}). We once again solve these by expanding the factors of $\lambda'$ on the left-hand side and matching coefficients to obtain $\nc{\mathfrak{e}}{n}{m}=\nc{\mathfrak{e}}{n}{m}(r)$. We then re-solve the equations, using now the four dimensional solution in terms of the coefficients $\mathfrak{b}$ in a general scheme change, noting now that contributions from $\mathfrak{e}$ will produce $\varepsilon$ terms in (\ref{eq:epsiloneqn}) to all orders in $\lambda$. We again find 36 constraints through six loops, which reduce to the 36 we had found previously. Thus, the vanishing of the constraints in $\overline{\text{MS}}$ implies the vanishing of the constraints in all schemes, so that the extension of the four dimensional solution to $d=4-\varepsilon$ is physically meaningful.

\section{Conclusion}
While the conclusion that the RG flow is gradient to six loops in $d=4$ and $d=4-\varepsilon$ if the conditions in \eqref{eq:Sconstraints} are satisfied is scheme-independent and thus physical, the specific $A$ and $G_{IJ}$ that we determine along the way are scheme-dependent. This means that they cannot be given a direct physical meaning along an RG flow. It would be interesting to understand the extent to which $A$ and $G_{IJ}$ may contain physical information, along and at a fixed point of the RG flow.

There are other cases where the gradient property of the RG flow has been examined in perturbation theory, e.g.\ \cite{Jack:2014pua, Jack:2015tka, Gracey:2015fia, Jack:2016utw}. In all these cases the RG flow is gradient. In cases with defects, the next-to-leading order result for the line-defect beta function in scalar-fermion theories in $d=4-\varepsilon$~\cite{Pannell:2023pwz} is a gradient vector field. We expect that a direct examination of the diagrammatic expansion for the gradient flow equation, analogous to the one presented here, will provide a perturbative proof of the gradient property in these line defect systems, given the relatively simple structure of the terms involved.

In summary, the evidence for the gradient property of RG flow from perturbation theory is overwhelming. It would be interesting to look beyond perturbation theory and seek connections with existing literature, e.g.\ \cite{Cotler:2022fze}, in specific models. Let us finally emphasise that the gradient property of RG flow cannot be viewed as a fundamental physical requirement at this stage, so it is important to explore how it may arise from deeper physical principles.
\medskip

\section*{Acknowledgements}
We would like to thank Ian Jack and especially Hugh Osborn for useful discussions and communications. We would also like to thank Slava Rychkov for discussions that inspired this work during ``Bootstrap 2023'' at ICTP-SAIFR in July 2023, and acknowledge support received through FAPESP grant 2021/14335-0. AS is supported by the Royal Society under grant URF{\textbackslash}R1{\textbackslash}211417 and by STFC under grant ST/X000753/1.

\appendix

\section{Explicit solution through two loops}\label{ExplicitAppendix}

In this appendix we exhibit explicitly the $\varepsilon$ corrections which arise in $A$ and $G^{IJ}$ consistent with (\ref{eq:epsiloneqn}). In the interest of space we will only display the solution through two-loop order. For $d=4$ this requires a solution only for the terms displayed in (\ref{eq:A1diag}), (\ref{eq:A2loops}) and (\ref{eq:G2loops}), but in $d=4-\varepsilon$ the additional contribution to $A$ in (\ref{eq:epsilonA}) requires the inclusion of terms from $\nc{G}{3}{}$. In the interest of exhibiting a simple solution, we set the coefficients which remain unfixed in this solution to zero. Absorbing the contributions of the $d$'s and $h$'s of \eqref{eq:additionalterm} into the corresponding $a$'s and $g$'s in the obvious way, we arrive at the following solution for the coefficients:
\begin{align}\label{eq:explicitsol}
    \nc{a}{1}{}&=1-\varepsilon\,,\nonumber\\
    \nc{a}{2}{1}&=\frac{1}{2304}\big(192 + 528 \varepsilon + 636 \varepsilon^2 + (1051 - 864\zeta_3) \varepsilon^3\big),\nonumber\\
    \nc{a}{2}{2}&=\frac{1}{32}\big(24 - 18 \varepsilon + (91 - 84 \zeta_3)\varepsilon^2\big) \,, \nonumber\\
    \nc{a}{2}{3}&=-\frac{1}{160}\big(420 - (760 - 
      480 \zeta_3) \varepsilon\nonumber\\ &\hspace{2cm} + (940 - 3 \pi^4 - 540 \zeta_3) \varepsilon^2\big)\varepsilon\,,\nonumber\\ \nc{g}{2}{}&=-3\,,\qquad \nc{g}{3}{1}=\nc{g}{3}{2}=0\,,\nonumber\\ \nc{g}{3}{3}&=\frac{1}{72} (24 - 21 \varepsilon - 4 \varepsilon^2)\,,\nonumber\\ \nc{g}{3}{4}&=\frac{1}{4} (6 - \varepsilon) \varepsilon\,, \nonumber\\
\nc{g}{3}{5}&=\frac{1}{8} \big(6 + (91 - 84 \zeta_3) \varepsilon\big)\,, \nonumber\\ \nc{g}{3}{6}&= -\frac{1}{40}\big(180 - (700 - 480 \zeta_3) \varepsilon\nonumber\\&\hspace{2cm}+ 3 (310 - \pi^4 - 180 \zeta_3) \varepsilon^2\big)\,, \nonumber\\ \nc{g}{3}{7}&=\frac{1}{192}\big(112+268\varepsilon+(361-288\zeta_3)\varepsilon^2\big)\,. 
\end{align}
With the results in \eqref{eq:explicitsol}, the solution may be represented using the graphical notation for tensor structures as
\begin{equation}
\begin{aligned}
    A &= -\frac{\varepsilon}{2}\,\begin{tikzpicture}[scale=0.5,baseline=(vert_cent.base)]
        \node (vert_cent) at (0,0) {$\phantom{\cdot}$};
        \draw (0,0) circle (1cm);
        \filldraw (-1,0) circle (2pt);
        \filldraw (1,0) circle (2pt);
        \draw (1,0) to[out=135, in=45] (-1,0);
        \draw (1,0) to[out=225, in=315] (-1,0);
    \end{tikzpicture}+\nc{a}{1}{}\,\begin{tikzpicture}[scale=0.5,baseline=(vert_cent.base)]
        \node (vert_cent) at (0,0) {$\phantom{\cdot}$};
        \node (center) at (0,0) {};
        \def\radius{1cm};
        \node[inner sep=0pt] (top) at (90:\radius) {};
        \node[inner sep=0pt] (left) at (210:\radius) {};
        \node[inner sep=0pt] (right) at (330:\radius) {};
        \draw (center) circle[radius=\radius];
        \filldraw (top) circle[radius=2pt];
        \filldraw (left) circle[radius=2pt];
        \filldraw (right) circle[radius=2pt];
        \draw (top) to[out=270, in=30] (left);
        \draw (top) to[out=270, in=150] (right);
        \draw (left) to[out=30, in=150] (right);
    \end{tikzpicture}+\nc{a}{2}{1}\,
    \begin{tikzpicture}[scale=0.5,baseline=(vert_cent.base)]
      \node (center) at (0,0) {};
      \def\radius{1cm};
      \node[inner sep=0pt] (n1) at (45:\radius) {};
      \node[inner sep=0pt] (n2) at (135:\radius) {};
      \node[inner sep=0pt] (n3) at (225:\radius) {};
      \node[inner sep=0pt] (n4) at (315:\radius) {};
      \draw (center) circle[radius=\radius];
      \filldraw (n1) circle[radius=2pt];
      \filldraw (n2) circle[radius=2pt];
      \filldraw (n3) circle[radius=2pt];
      \filldraw (n4) circle[radius=2pt];
      \draw (n1) to[out=225, in=135] (n4);
      \draw (n1) -- (n4);
      \draw (n2) to[out=315, in=45] (n3);
      \draw (n2) -- (n3);
    \end{tikzpicture}
  \\&\quad+ \nc{a}{2}{2}\,
  \begin{tikzpicture}[scale=0.5,baseline=(vert_cent.base)]
      \node (center) at (0,0) {};
      \def\radius{1cm};
      \node[inner sep=0pt] (n1) at (0:\radius) {};
      \node[inner sep=0pt] (n2) at (90:\radius) {};
      \node[inner sep=0pt] (n3) at (180:\radius) {};
      \node[inner sep=0pt] (n4) at (270:\radius) {};
      \draw (center) circle[radius=\radius];
      \filldraw (n1) circle[radius=2pt];
      \filldraw (n2) circle[radius=2pt];
      \filldraw (n3) circle[radius=2pt];
      \filldraw (n4) circle[radius=2pt];
      \draw (n1) to[out=180, in=270] (n2);
      \draw (n2) to[out=270, in=0] (n3);
      \draw (n3) to[out=0, in=90] (n4);
      \draw (n4) to[out=90, in=180] (n1);
    \end{tikzpicture}
  + \nc{a}{2}{3}\,
  \begin{tikzpicture}[scale=0.5,baseline=(vert_cent.base)]
      \node (center) at (0,0) {};
      \def\radius{1cm};
      \node[inner sep=0pt] (n1) at (0:\radius) {};
      \node[inner sep=0pt] (n2) at (90:\radius) {};
      \node[inner sep=0pt] (n3) at (180:\radius) {};
      \node[inner sep=0pt] (n4) at (270:\radius) {};
      \draw (center) circle[radius=\radius];
      \filldraw (n1) circle[radius=2pt];
      \filldraw (n2) circle[radius=2pt];
      \filldraw (n3) circle[radius=2pt];
      \filldraw (n4) circle[radius=2pt];
      \draw (n1) to[out=150, in=30] (n3);
      \draw (n1) to[out=210, in=-30] (n3);
      \draw (n2) to[out=300, in=60] (n4);
      \draw (n2) to[out=240, in=120] (n4);
    \end{tikzpicture}+\text{O}(\lambda^5)\,,
\end{aligned}
\end{equation}

\begin{equation}
\begin{aligned}
    G^{IJ}&=\begin{tikzpicture}[xscale=0.6,yscale=0.5,
    vertex/.style={draw,circle,fill=black,minimum size=2pt,inner sep=0pt},
    arc/.style={},baseline=(vert_cent.base)]
    \node (vert_cent) at (0,-1.25) {$\phantom{\cdot}$};
    \foreach [count=\i] \coord in {
(1.00,0), (-1,0),(-1,-0.75),(1,-0.75)}{
        \node[] (p\i) at \coord {};
    }
    \foreach [count=\i] \coord in {
(-1,-1.5), (1,-1.5),(-1,-2.25),(1,-2.25)}{
        \node[] (d\i) at \coord {};
    }
    \draw[arc] (p1) edge (p2);
    \draw[arc] (p4) edge (p3);
    \draw[arc] (d1) edge (d2);
    \draw[arc] (d3) edge (d4);
\end{tikzpicture}
+\nc{g}{2}{}\begin{tikzpicture}[xscale=0.6,yscale=0.5,
    vertex/.style={draw,circle,fill=black,minimum size=2pt,inner sep=0pt},
    arc/.style={},baseline=(vert_cent.base)]
    \node (vert_cent) at (0,-1.25) {$\phantom{\cdot}$};
    \node[vertex] (c) at (0,-0.375) {};
    \foreach [count=\i] \coord in {
(1.00,0), (-1,0),(-1,-0.75),(1,-0.75)}{
        \node[] (p\i) at \coord {};
    }
    \foreach [count=\i] \coord in {
(-1,-1.5), (1,-1.5),(-1,-2.25),(1,-2.25)}{
        \node[] (d\i) at \coord {};
    }
    \draw[arc] (c) edge (p1)
                   edge (p2)
                   edge (p3)
                   edge (p4);
    \draw[arc] (d1) edge (d2);
    \draw[arc] (d3) edge (d4);
\end{tikzpicture} +\nc{g}{3}{1}\begin{tikzpicture}[xscale=0.6,yscale=0.5,
    vertex/.style={draw,circle,fill=black,minimum size=2pt,inner sep=0pt},
    arc/.style={},baseline=(vert_cent.base)]
    \node (vert_cent) at (0,-1.25) {$\phantom{\cdot}$};
    \node[vertex] (c1) at (0,-0.755) {};
    \node[vertex] (c2) at (0,-1.75) {};
    \foreach [count=\i] \coord in {
(1.00,-0.75), (-1,0),(-1,-0.75),(1,-1)}{
        \node[] (p\i) at \coord {};
    }
    \foreach [count=\i] \coord in {
(-1,-1.5), (1,-1.75),(-1,-1.75),(1,-2.5)}{
        \node[] (d\i) at \coord {};
    }
    \draw[arc] (c1) edge (p1)
                   edge (p2)
                   edge (p3)
                   edge (d1);
    \draw[arc] (c2) edge (p4)
                   edge (d2)
                   edge (d3)
                   edge (d4);
\end{tikzpicture}\\&\quad+\nc{g}{3}{2}\begin{tikzpicture}[xscale=0.6,yscale=0.5,
    vertex/.style={draw,circle,fill=black,minimum size=2pt,inner sep=0pt},
    arc/.style={},baseline=(vert_cent.base)]
    \node (vert_cent) at (0,-1.25) {$\phantom{\cdot}$};
    \node[vertex] (c1) at (0,-0.5) {};
    \node[vertex] (c2) at (0,-1.75) {};
    \foreach [count=\i] \coord in {
(1.00,0), (-1,0),(-1,-1),(1,-1)}{
        \node[] (p\i) at \coord {};
    }
    \foreach [count=\i] \coord in {
(-1,-1.25), (1,-1.25),(-1,-2.25),(1,-2.25)}{
        \node[] (d\i) at \coord {};
    }
    \draw[arc] (c1) edge (p1)
                   edge (p2)
                   edge (p3)
                   edge (p4);
    \draw[arc] (c2) edge (d1)
                   edge (d2)
                   edge (d3)
                   edge (d4);
\end{tikzpicture} + \nc{g}{3}{3}\begin{tikzpicture}[xscale=0.6,yscale=0.5,
    vertex/.style={draw,circle,fill=black,minimum size=2pt,inner sep=0pt},
    arc/.style={},baseline=(vert_cent.base)]
    \node (vert_cent) at (0,-1.25) {$\phantom{\cdot}$};
    \node[vertex] (c1) at (-0.25,-0.75) {};
    \node[vertex] (c2) at (0.25,-0.75) {};
    \foreach [count=\i] \coord in {(-1,0),(-1,-0.75),(-1,-1.5)}{
        \node[] (p\i) at \coord {};
    }
    \foreach [count=\i] \coord in {(1,0),(1,-0.75),(1,-1.5)}{
        \node[] (d\i) at \coord {};
    }
    \node[] (d4) at (1,-2) {};
    \node[] (p4) at (-1,-2) {};
    \draw[arc] (c1) edge (p1)
                   edge (p2)
                   edge (p3)
                   edge (c2);
    \draw[arc] (c2) edge (d1)
                   edge (d2)
                   edge (d3);
    \draw[arc] (p4) edge (d4);
\end{tikzpicture}
\\&\quad+\nc{g}{3}{4}\begin{tikzpicture}[xscale=0.6,yscale=0.5,
    vertex/.style={draw,circle,fill=black,minimum size=2pt,inner sep=0pt},
    arc/.style={},baseline=(vert_cent.base)]
    \node (vert_cent) at (0,-1.25) {$\phantom{\cdot}$};
    \node[vertex] (c1) at (0,-0.375) {};
    \node[vertex] (c2) at (0,-1.125) {};
    \foreach [count=\i] \coord in {(-1,0),(-1,-0.75),(-1,-1.5)}{
        \node[] (p\i) at \coord {};
    }
    \foreach [count=\i] \coord in {(1,0),(1,-0.75),(1,-1.5)}{
        \node[] (d\i) at \coord {};
    }
    \node[] (d4) at (1,-2) {};
    \node[] (p4) at (-1,-2) {};
    \draw[arc] (c1) edge (p1)
                   edge (p2)
                   edge (d1)
                   edge (c2);
    \draw[arc] (c2) edge (p3)
                   edge (d2)
                   edge (d3);
    \draw[arc] (p4) edge (d4);
\end{tikzpicture}
+\nc{g}{3}{5}\begin{tikzpicture}[xscale=0.6,yscale=0.5,
    vertex/.style={draw,circle,fill=black,minimum size=2pt,inner sep=0pt},
    arc/.style={},baseline=(vert_cent.base)]
    \node (vert_cent) at (0,-1.25) {$\phantom{\cdot}$};
    \node[vertex] (c1) at (-0.25,-0.375) {};
    \node[vertex] (c2) at (0.25,-0.375) {};
    \foreach [count=\i] \coord in {(-1,0),(-1,-0.75),(-1,-1.25)}{
        \node[] (p\i) at \coord {};
    }
    \foreach [count=\i] \coord in {(1,0),(1,-0.75),(1,-1.25)}{
        \node[] (d\i) at \coord {};
    }
    \node[] (d4) at (1,-2) {};
    \node[] (p4) at (-1,-2) {};
    \draw[arc] (c1) edge (p1)
                   edge (p2)
                   edge[bend right=30] (c2)
                   edge[bend left=30] (c2);
    \draw[arc] (c2) edge (d1)
                   edge (d2);
    \draw[arc] (p3) edge (d3);
    \draw[arc] (p4) edge (d4);
\end{tikzpicture} 
\\&\quad+\nc{g}{3}{6}\begin{tikzpicture}[xscale=0.6,yscale=0.5,
    vertex/.style={draw,circle,fill=black,minimum size=2pt,inner sep=0pt},
    arc/.style={},baseline=(vert_cent.base)]
    \node (vert_cent) at (0,-1.25) {$\phantom{\cdot}$};
    \node[vertex] (c1) at (0,0) {};
    \node[vertex] (c2) at (0,-0.75) {};
    \foreach [count=\i] \coord in {(-1,0),(-1,-0.75),(-1,-1.5)}{
        \node[] (p\i) at \coord {};
    }
    \foreach [count=\i] \coord in {(1,0),(1,-0.75),(1,-1.5)}{
        \node[] (d\i) at \coord {};
    }
    \node[] (d4) at (1,-2.25) {};
    \node[] (p4) at (-1,-2.25) {};
    \draw[arc] (c1) edge (p1)
                   edge (d1)
                   edge[bend right=30] (c2)
                   edge[bend left=30] (c2);
    \draw[arc] (c2) edge (p2)
                   edge (d2);
    \draw[arc] (p3) edge (d3);
    \draw[arc] (p4) edge (d4);
\end{tikzpicture}
+\nc{g}{3}{7}\begin{tikzpicture}[xscale=0.6,yscale=0.5,
    vertex/.style={draw,circle,fill=black,minimum size=2pt,inner sep=0pt},
    arc/.style={},baseline=(vert_cent.base)]
    \node (vert_cent) at (0,-1.25) {$\phantom{\cdot}$};
    \node[vertex] (c1) at (-0.5,0) {};
    \node[vertex] (c2) at (0.5,0) {};
    \foreach [count=\i] \coord in {(-1,0),(-1,-0.75),(-1,-1.5)}{
        \node[] (p\i) at \coord {};
    }
    \foreach [count=\i] \coord in {(1,0),(1,-0.75),(1,-1.5)}{
        \node[] (d\i) at \coord {};
    }
    \node[] (d4) at (1,-2.25) {};
    \node[] (p4) at (-1,-2.25) {};
    \draw[arc] (c1) edge (p1)
                   edge (c2)
                   edge[bend right=30] (c2)
                   edge[bend left=30] (c2);
    \draw[arc] (c2) edge (d1);
    \draw[arc] (p2) edge (d2);
    \draw[arc] (p3) edge (d3);
    \draw[arc] (p4) edge (d4);
\end{tikzpicture}+\text{O}(\lambda^3)\,,
\end{aligned}
\end{equation}
where it is understood that each term in the metric includes a sum over permutations of indices on each side. As one can check, this solution for $G^{IJ}$ and $A$ correctly reproduces the terms up to $\text{O}(\lambda^2)$ in the beta function
\begin{equation}
\begin{aligned}
    \beta&=-\varepsilon\,\begin{tikzpicture}[scale=0.5,baseline=(vert_cent.base)]
        \node (vert_cent) at (0,0) {$\phantom{\cdot}$};
        \filldraw (0,0) circle (2pt);
        \draw (-1,1)--(0,0);
        \draw (-1,-1)--(0,0);
        \draw (1,1)--(0,0);
        \draw (1,-1)--(0,0);
    \end{tikzpicture}+\begin{tikzpicture}[scale=0.5,baseline=(vert_cent.base)]
        \node (vert_cent) at (0,0) {$\phantom{\cdot}$};
        \draw (0,0) circle (1cm);
        \filldraw (-1,0) circle (2pt);
        \filldraw (1,0) circle (2pt);
        \draw (-1.7,0.7)--(-1,0);
        \draw (-1.7,-0.7)--(-1,0);
        \draw (1.7,0.7)--(1,0);
        \draw (1.7,-0.7)--(1,0);
    \end{tikzpicture}-\begin{tikzpicture}[scale=0.5,baseline=(vert_cent.base)]
      \node (vert_cent) at (0,0) {$\phantom{\cdot}$};
      \draw (-1,-1)--(1,1);
      \draw (-1,1)--(1,-1);
      \filldraw (0,0) circle[radius=2pt];
      \filldraw (1,1) circle[radius=2pt];
      \filldraw (1,-1) circle[radius=2pt];
      \draw (1,1)--(2,1);
      \draw (1,-1)--(2,-1);
      \draw (1,1) to[out=240, in=120] (1,-1);
      \draw (1,1) to[out=300, in=60] (1,-1);
    \end{tikzpicture}\\&\quad
    + \frac{1}{12}\,
    \begin{tikzpicture}[scale=0.5,baseline=(vert_cent.base)]
      \draw (-1,0)--(1,0);
      \draw (-0.8,1)--(0,0);
      \draw (-0.8,-1)--(0,0);
      \filldraw (0,0) circle[radius=2pt];
      \filldraw (0.7,0) circle[radius=2pt];
      \filldraw (2.3,0) circle[radius=2pt];
      \draw (1,0)--(3,0);
      \draw (0.7,0) to[out=60, in=120] (2.3,0);
      \draw (0.7,0) to[out=-60, in=-120] (2.3,0);
    \end{tikzpicture}+\text{O}(\lambda^3)\,,
\end{aligned}
\end{equation}
where again there is understood to be a sum over distinct permutations of the free indices in each term. In particular, one can verify that (\ref{eq:explicitsol}) satisfies (\ref{eq:2loopeq1}), (\ref{eq:2loopeq2}) and (\ref{eq:2loopeq3}) up to $\text{O}(\varepsilon)$ terms. One should note that this is but a single allowable solution, and that the freedom contained in the unfixed coefficients could be fixed by imposing conditions on $A$ and $G^{IJ}$, for instance that $A$ contains terms at most linear in $\varepsilon$.

\bibliographystyle{elsarticle-num}
\bibliography{main}

\end{document}